\documentclass[11pt]{article} 
\usepackage[margin=1in]{geometry}
\usepackage{amsmath, amsthm, amsfonts}

\usepackage{graphicx}
\usepackage{tikz}
 \usepackage{pgfplots}
 \usepackage{pgf-pie}
\pgfplotsset{compat=1.5.1}

\usepackage{graphicx}
\graphicspath{ {./images2/} }
\usepackage{subcaption}
\usepackage{url}
\usepackage{booktabs}

\providecommand{\keywords}[1]
{
  \small	
  \textbf{\textit{Keywords---}} #1
}

\theoremstyle{definition}

\theoremstyle{remark}

\title{Usage Analysis of Mobile Devices}
\author{ 
Aman Singh \\ \small IIT Roorkee
\and Ashish Prajapatia \\ \small IIT Roorkee
\and VikashKumar \\ \small IIT Roorkee
\and 
Subhankar Mishra  \\
\small School of Computer Sciences \\
\small National Institute of Science Education and Research\\
\small Bhubaneswar-752050, Odisha, India \\
\small Homi Bhabha National Institute \\
\small Anushaktinagar, Mumbai - 400094, India\\
\small smishra@niser.ac.in
}
\date{}
\begin{document}
\maketitle

\abstract{
  Mobile devices have evolved from just communication devices into an indispensable part of people's lives in form of smartphones, tablets and smart watches. Devices are now more personal than ever and carry more information about a person than any other. Extracting user behaviour is rather difficult and time-consuming as most of the work previously has been manual or requires feature extraction. In this paper, a novel approach of user behavior detection is proposed with Deep Learning Network (DNN). Initial approach was to use recurrent neural network (RNN) along with LSTM for completely unsupervised analysis of mobile devices. Next approach is to extract features by using Long Short Term Memory (LSTM) to understand the user behaviour, which are then fed into the Convolution Neural Network (CNN). This work mainly  concentrates on detection of user behaviour and anomaly detection for usage analysis of mobile devices. Both the approaches are compared against some baseline methods. Experiments are conducted on the publicly available dataset to show that these methods can successfully capture the user behaviors. 
}
\bigskip

\section{Introduction}

Mobile devices have become extremely important in the last few years but little public information exists on mobile application usage behavior. Today, there are more than 2.8 million apps available for the Android platform and 2.2 million for Apple’s iPhone.

Despite these large numbers, there is little public research available on application usage behavior. Very basic questions remain unanswered \cite{Bohmer2011,vallina2010,Tong2013}. For instance, how long does each interaction with an app last? Does this vary by application category? If so, which categories inspire the longest interactions with their users? The data on context’s effect on application usage is equally sparse, leading to additional interesting questions. Which app user uses the most? How many apps a users generally uses per day? How many users are using the particular app on a particular day.

These data can be very useful to the Mobile Application Developers who will be able to think of great apps in the categories which are famous for the regular users and implement features that the users need in their app. This survey also helps in understanding where the current youth in an academic institutions spends his/her most of the time on the mobile device.

\section{Method}

We started by creating the application in Android Studio and choosing what
data should we collect from the users. The collected data should be such that we get some promising statistics at the end of the project. We decided to collect foreground usage time for various apps in the user’s phone and the network used by the user. Also, we had to start a server to collect that data and save it to a database. So, we collected data in these three forms:

\begin{enumerate}
	\item InstanceInfo (appID, state, profession, gender, age)
	\item ForegroundTime (appID, packageName, foregroundTime)
	\item HourlyData (appID, packageName, foregroundTime, network)
\end{enumerate}

In the following subsections, we briefly describe the tools used, android application and the server side operation. 

\subsection{Tools Used}
For the project, we have used a number of tools to complete it. They are as follows: 
\begin{enumerate}
	\item Android Studio \cite{android} : Android Studio is the official integrated development environment (IDE) for the Android platform. We used it to create the application and a test android phone to test the app.
	\item DigitalOcean \cite{DigitalOcean} : DigitalOcean provides developers cloud services that help to deploy and scale applications that run simultaneously on multiple computers. We bought a server on this platform to install our server and host our api to receive the data over the internet.
	\item Apache HTTP Server \cite{apache} : The Apache HTTP Server, colloquially called Apache, is the
	world's most used web server software. We used this to host the api to save data on the server. Also, we used it to host a web page to give a link to download the apk of the android application.
	\item MySQL \cite{MySQL} : MySQL is an open-source relational database management system (RDBMS). We installed it on the server and used it to store the data in a tabular form. We created three tables namely, InstanceInfo, ForegroundTime and HourlyData to store the three types of data sent by the application.
	\item PhpMyAdmin \cite{php}: phpMyAdmin is a free and open source tool written in PHP intended to handle the administration of MySQL or MariaDB with the use of a web browser. We used it to access and monitor our data. Also, we used the sql queries in it to access the required data. It also helped us to generate the graphs in this report.
	\item Python \cite{python} : Python is a widely used high-level programming language for general-purpose programming, created by Guido van Rossum and first released in 1991. When we got data from the UsageStats, it had package name of the app and we needed the app category for some statistics. So, we used python to scrape data from the google play website to get the app category.
\end{enumerate}

\subsection{Android Application}

On starting the app, it asks the user to fill some information about him like gender, state and profession. No private information is asked that he is unwilling to share to a unknown guy. Until, the user fills in the info, the user is unable to proceed with the application and no data is collected from the user. Once the user fills in the his information, it sends the data to the server to save. To display on the screen, the app collects foreground time of the applications used by the users for the last 24 hrs and display it on the screen. On clicking any element on the table, the app shows how many users have used the app in the last 6 hrs, 12 hrs and 24 hrs of time. \\ 
In background, the app securely sends the collected data to the server, at every hour after the time of filling in of the survey page. The data sent in JSON form which contains the data about the application used, their foreground time and the network the user was working on. This data is sent once every hour. Another data that was being sent by the phone was the same data but for an interval of a day and was sent once every day. \\ 
The application was made so that no information that the user is not willing to share with anyone is never collected by the application. To convince the users, we had to put up the full code of the application on a public platform, Github, so that the users can take a look if they ever doubt us.

The application, named 'AppUsageStats', is an application for Android version Marshmallow or above. The main feature of the app was to access the foreground time for other applications. To access this, we needed to use the class UsageStats and UsageStatsManager in Android Studio. Due to these classes we had to set our base android version to Marshmallow. To access these classes, we asked the user for the permission ‘android.permission.PACKAGE\_USAGE\_STATS’ which the user permit by going to setting and allowing the app to use it.

To send the data over internet, we needed OS to give us the access to the available Internet for which we asked the permission for 'android.permission. INTERNET'. Also, we used Volley to send requests to the server. We also checked for the network being used with the help ConnectivityManager class available in Android. We were able to check whether mobile is on 2G, 3G, 4G, or Wifi network.

To send the data in background, we first used JobScheduler class to call our background service at regular intervals. But, for some reasons it was not being called when the phone was in locked state. We tried debugging it, but the result was the same. We gave it the required permission 'android.permission. WAKE\_LOCK', but nothing changed. So, we shifted to AlarmManager.
To send the data in background, we needed the same permission as for JobScheduler. We used AlarmManager class to call our background service that sent the data to the server. It called the background service around the same time as was needed. But, it was not accurate in time. This actually came to be good for us as now different phones were not sending the data at the same time.

\subsection{Server}

The storage server was set up after buying a server on DigitalOcean and starting a apache server on it. When the application sends a POST request from the phone, the server handles the requests using the PHP code on it. The PHP code saves the data to a database on the same server. It had different PHP files for different data sent by the phones. So, the server hosted three PHP files namely, instanceInfo.php, foregroundTime.php and hourlyData.php and getHourlyData.php. The instanceInfo.php file handled requests that are sent when the app user completes the survey form on the app. The foregroundTime.php handled the daily data sent by the app once a day. The hourlyData.php handled data sent by the app once every hour. The getHourlyData.php was used to get data from the server about how many users have used an app in the given hour.

\section{Results and Discussion}

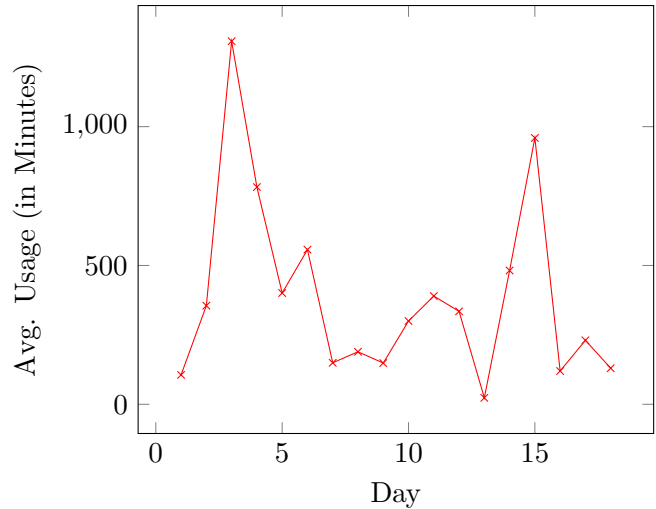
\begin{figure}[h]
		\centering
	\begin{tikzpicture}
	    \begin{axis}[
	        xlabel=Day,
	        ylabel=Avg. Usage (in Minutes)]
	    \addplot[color=red,mark=x] coordinates {
	        (1,105.8177)
	(2,355.17401111)
	(3,1307.80543333)
	(4,782.4001)
	(5,400.7865)
	(6,556.93502)
	(7,149.38144)
	(8,189.13514)
	(9,148.44112)
	(10,299.9036)
	(11,389.78532)
	(12,334.6616)
	(13,23.18665)
	(14,481.8021)
	(15,959.3302)
	(16,119.881)
	(17,230.32965)
	(18,130.0108)
	};
	    \end{axis}
	    
	\end{tikzpicture}
	\caption{Shows the average usage of apps per day on the given dates.}
	\label{fig:avgusageday}
	\end{figure}

The data was collected from 32 users for a period of 18 days. In Fig \ref{fig:avgusageday}, the average usage of application per day has been shown. The days were selected continuously and carefully to avoid any external factors such as spring fest or sports meet in order to have a normal dataset rather than skewed one. However getting such a long period of contiguous days was not easy and we hindered by the industrial trip week; where the students attended for their respective events which has lead to different usage scenarios as seen in Fig \ref{fig:avgusageday}.

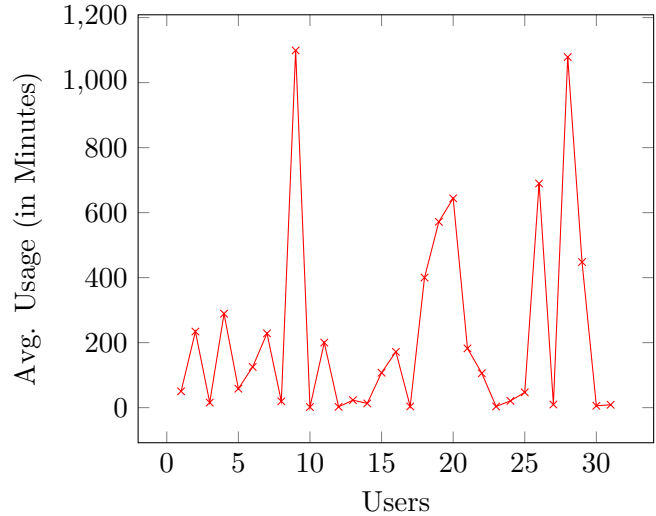
\begin{figure}[h]
	\centering
	\begin{tikzpicture}
	\begin{axis}[
	xlabel=Users,
	ylabel=Avg. Usage (in Minutes)]
	\addplot[color=red,mark=x] coordinates {
		(1,50.2787)
(2,234.1473)
(3,15.6362)
(4,288.9384)
(5,58.2781)
(6,125.3363)
(7,228.5507)
(8,19.5208)
(9,1099.003)
(10,1.77373333)
(11,200.1615)
(12,2.5197)
(13,23.0659)
(14,13.06325)
(15,107.47475)
(16,171.7353)
(17,3.796)
(18,400.2579)
(19,571.78335)
(20,643.956925)
(21,182.31089)
(22,105.8177)
(23,3.6261)
(24,20.8041)
(25,47.0298)
(26,689.64242308)
(27,9.69995)
(28,1078.668)
(29,447.94113333)
(30,5.7003)
(31,8.91665)
	};
	\end{axis}
	
	\end{tikzpicture}
	\caption{Shows the average time used by each id through the whole process.}
	\label{fig:avgusageuser}
\end{figure}

In Fig \ref{fig:avgusageuser}, we show the  average app usage by various users (tagged by their user IDs to avoid linking to the original user). Some skeptical users did not follow through with letting the usage of our app in order to collect data and hence the lower avg usage times on their devices.

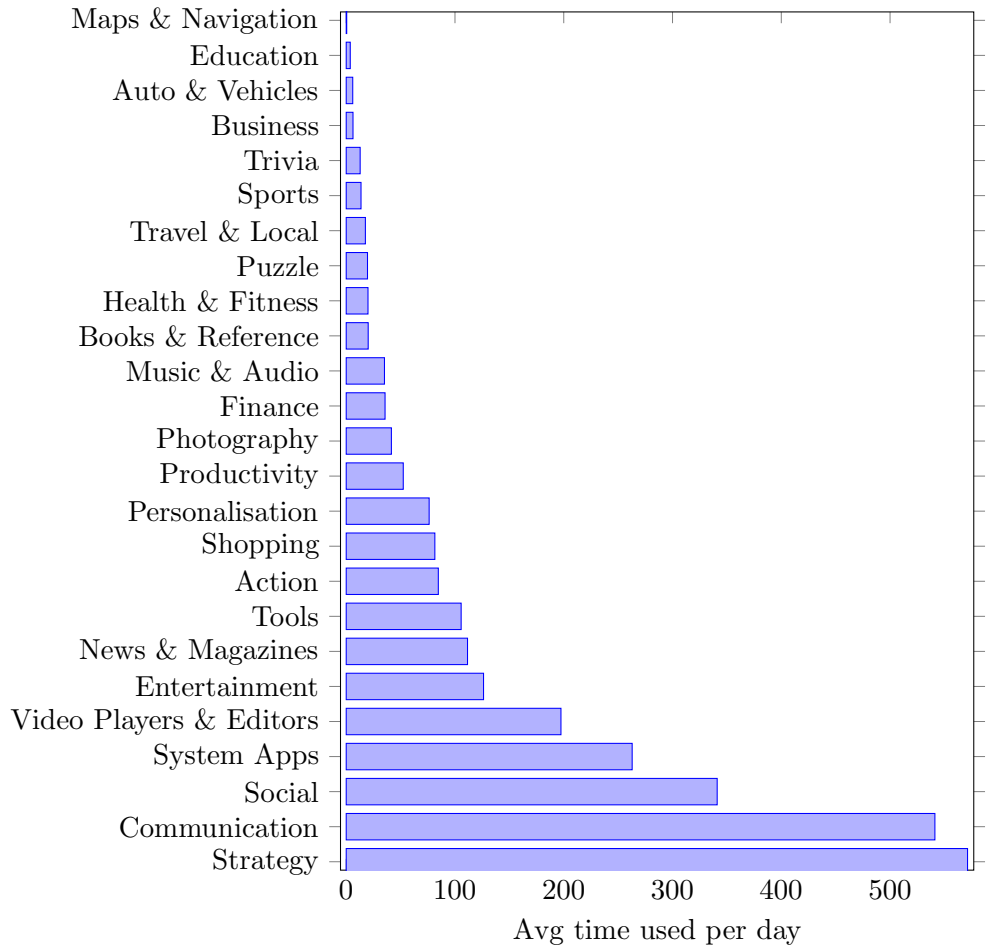
\begin{figure}[ht]
	\centering
\begin{tikzpicture}
  \begin{axis}[
    xbar,
    width=10cm, 
    height=13cm, 
    enlargelimits=0.01,
    xlabel={Avg time used per day},
    symbolic y coords={Strategy,Communication,Social,System Apps,Video Players \& Editors,Entertainment, News \& Magazines, Tools, Action, Shopping, Personalisation,Productivity,Photography, Finance, Music \& Audio, Books \& Reference,Health \& Fitness,Puzzle,Travel \& Local,Sports,Trivia,Business,Auto \& Vehicles,Education, Maps \& Navigation},
    ytick=data,
    ]
    \addplot coordinates {
    (571.25,Strategy)
(541.23,Communication)
(341.06,Social)
(262.9,System Apps)
(197.43,Video Players \& Editors)
(126.26,Entertainment)
(111.53,News \& Magazines)
(105.67,Tools)
(84.68,Action)
(81.49,Shopping)
(76.25,Personalisation)
(52.45,Productivity)
(41.54,Photography)
(35.71,Finance)
(35.17,Music \& Audio)
(20.21,Books \& Reference)
(20.03,Health \& Fitness)
(19.57,Puzzle)
(17.63,Travel \& Local)
(13.61,Sports)
(12.82,Trivia)
(6.24,Business)
(6,Auto \& Vehicles)
(3.72,Education)
(0.46,Maps \& Navigation)
};
  \end{axis}
\end{tikzpicture}
	\caption{Shows the average time used by each id through the whole process.}
	\label{fig:avgusageapp}
\end{figure}

Next, we analyze the category of apps (obtained from Google Play Store) that has been used by the users during the whole process. As we can see in Fig \ref{fig:avgusageapp},Strategy which is a category for a games (Game called Dominations was very popular during that time) was the most popular category, followed by Communication and Social that includes Facebook, Facebook Messenger, Whatsapp among others. 

\begin{figure}[h]
	\centering
	\begin{tikzpicture} 
		\pie{71.7/Wi-Fi, 0.9/3G, 27.2/4G, 0.2/2G}
	\end{tikzpicture}
	\caption{Shows the percentage of network mode used by the users to connect to the Internet.}
	\label{fig:network}
\end{figure}
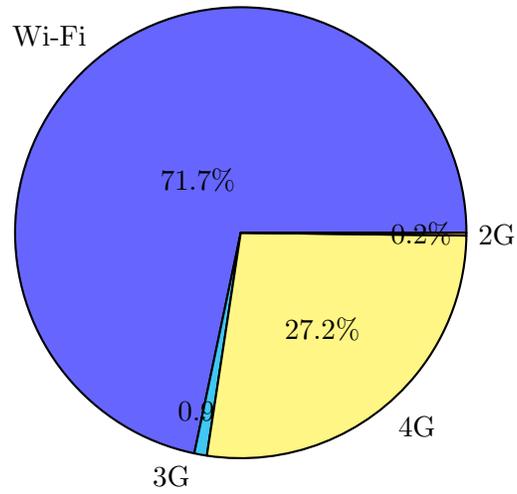

Given the Wi-Fi campus, the maximum usage derives from it. With the current addition of 4G phones and free 4G offerings, the 2nd highest usage has been attributed to it as seen in Fig \ref{fig:network}.

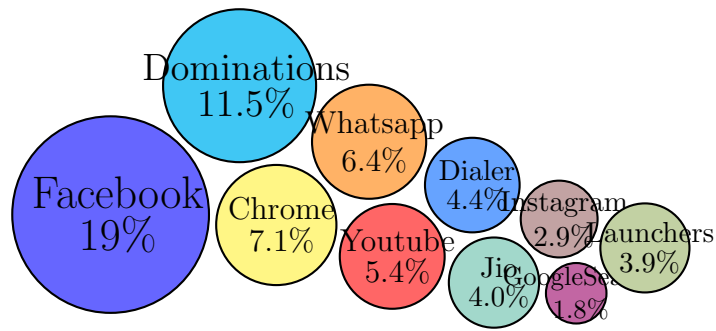
\begin{figure}[h]
	\centering
	\begin{tikzpicture} 
	\pie[cloud, text=inside, scale font]{19/Facebook, 11.5/Dominations, 7.1/Chrome, 6.4/Whatsapp, 5.4/Youtube, 4.4/Dialer, 4.0/Jio, 2.9/Instagram, 1.8/GoogleSearch, 3.9/Launchers}
	\end{tikzpicture}
	\caption{Shows the percentage usage of top 10 apps with respect to the total app usages.}
	\label{fig:10avgusageday}
\end{figure}

Last but not the least, we analyze the apps that have occupied the most time on screen of the users during those periods with Facebook dominating at $19\%$. Top 10 apps have been shown in Fig \ref{fig:10avgusageday}. Social. communication and games have occupied most of the time on devices which could have been intuitively predicted and also match to the observations from \cite{jacob2014,saxena2014}

\section{Conclusion }

In this paper, we have presented the usage analysis of mobile phones of 32 students over a period of 18 days. In short, this paper included the following contributions (amongst others): average usage time differs extensively between app categories, • a context-related analysis that led to the following conclusions (among other findings): (1) mobile phones are still used mostly for communication (text and voice); (2) some apps have somewhat intense spikes in relative usage (e.g. music and social apps), whereas others are more generally average used; (3) when people actively use their devices they spend less time with each app. (4) Most app used by males and females.(5) Which network mode people use more for the Internet use.

As the mobile is becoming more personal than ever; it is much harder to get a large number of students who are well aware of the potential data collection issues. However, we believe that the data we collected along with the analysis will be very useful to understand the mobile usage trend among the current generation. We are working towards effectively increasing the scale and diversity of the survey to gain a better understanding the mobile device usage.

\section{Issues}

Although we find the overall survey and experiment to be successful, we did experience some issues which we would like to share. As the app was supposed to collect data from the users, most of the students around us were not willing to install the app. We did our best to convince them that the app was not collecting any private data, not any password or other things. We even put our whole code on Github to let the users check if the app was collecting any data that it was not supposed to. Hence the number of users is 32. Given the current mass malware and ransomware threats around the world and students being more informed and scared; it was hard to convince the users to include the application in their mobile phones for monitoring the usage. 

Restrictions on the newer versions of android lets us recover less information about mobile phone usage without root access as compared to older versions of android which most of the previous literature have covered \cite{jacob2014, saxena2014} which would have led to much quantitative as well as qualitative data. However root access again comes with its own vulnerability.

\end{document}